\documentclass[conference, 10pt]{IEEEtran}
\IEEEoverridecommandlockouts
\usepackage{cite}
\usepackage{amsmath,amssymb,amsfonts}
\usepackage{algorithmicx}
\usepackage{graphicx}
\usepackage{textcomp}
\usepackage{xcolor}
\usepackage{algorithm}
\usepackage[noend]{algpseudocode}
\usepackage{booktabs} % 导入booktabs包
\usepackage{subfigure}
\usepackage{amsmath}
\usepackage{xcolor}
\def\BibTeX{{\rm B\kern-.05em{\sc i\kern-.025em b}\kern-.08em
    T\kern-.1667em\lower.7ex\hbox{E}\kern-.125emX}}
\begin{document}

\title{Joint Model Assignment and Resource Allocation for Cost-Effective Mobile Generative Services}
%Edge-Enabled Model Assignment and Resource Allocation for AIGC Services
%Edge-Enabled Model and Resource Allocation for Enhanced AI-Generated Content (AIGC) Services
%Edge-Enabled Model Assignment and Resource Allocation for Enhanced AIGC Services
\author{\IEEEauthorblockN{Shuangwei~Gao\IEEEauthorrefmark{1},~Peng~Yang\IEEEauthorrefmark{1},~Yuxin~Kong\IEEEauthorrefmark{1},~Feng~Lyu\IEEEauthorrefmark{2}, and~Ning Zhang\IEEEauthorrefmark{3}}
	\IEEEauthorblockA{\IEEEauthorrefmark{1}School of Electronic Information and Communications, Huazhong University of Science and Technology, Wuhan, China \\
        \IEEEauthorrefmark{2}School of Computer Science and Engineering, Central South University, Changsha, China \\
        \IEEEauthorrefmark{3}Department of Electrical and Computer Engineering, University of Windsor, Windsor, ON, Canada \\
 }
	Email: \IEEEauthorrefmark{1}\{sw$\_$gao, yangpeng, yxkong\}@hust.edu.cn,
        \IEEEauthorrefmark{2}fenglyu@csu.edu.cn, 
        \IEEEauthorrefmark{3}ning.zhang@uwindsor.ca}
\maketitle

\begin{abstract}
Artificial Intelligence Generated Content (AIGC) services can efficiently satisfy user-specified content creation demands, but the high computational requirements pose various challenges to supporting mobile users at scale. In this paper, we present our design of an edge-enabled AIGC service provisioning system to properly assign computing tasks of generative models to edge servers, thereby improving overall user experience and reducing content generation latency. Specifically, once the edge server receives user requested task prompts, it dynamically assigns appropriate models and allocates computing resources based on features of each category of prompts. The generated contents are then delivered to users. The key to this system is a proposed probabilistic model assignment approach, which estimates the quality score of generated contents for each prompt based on category labels. Next, we introduce a heuristic algorithm that enables adaptive configuration of both generation steps and resource allocation, according to the various task requests received by each generative model on the edge.
%we introduce an algorithm that generates an approximate optimal solution based on task requests received by each generative model, enabling adaptive configuration of generation steps and resource allocation.
%we introduce an algorithm that facilitates adaptive configuration of both generation steps and resource allocation, according to the task requests received by each generative model on the edge.
Simulation results demonstrate that the designed system can effectively enhance the quality of generated content by up to 4.7\% while reducing response delay by up to 39.1\% compared to benchmarks.
\end{abstract}

\section{Introduction}

The Artificial Intelligence Generated Content (AIGC) technique is undergoing rapid advancement, driven by sophisticated generative AI models that can produce various contents, including text, images, and videos \cite{wahle2023ai}. Examples of these models include OpenAI's large multimodal model, ChatGPT4 \cite{achiam2023gpt}, image-centric diffusion models, DALL-E3 \cite{betker2023improving} and Stable diffusion \cite{rombach2022high}.
In terms of content generation and information visualization, the AIGC technique provides an efficient way to meet diverse user demands \cite{cetinic2022understanding}.
As AIGC pushes the boundary of customized content creation, it also faces significant challenges. The complexity of the models leads to high training costs and resource demands, making it difficult for users with limited resources to obtain quality contents in a cost-effective manner \cite{xu2024unleashing,huang2024joint}. For instance, the latest  video generation model, Sora \cite{sora}, requires 324 GFLOPs of computing power for training an image of 1024*768 pixels. Therefore, the challenge of high computing demands has become a critical concern in AIGC service provisioning, especially as the demand for imaginary and creative contents grows.

To facilitate the availability of the AIGC services and mitigate the challenge of intensive computing requirements, deploying AIGC models on edge servers has become a promising solution. Du \textit{et al.} proposed the concept of \textit{AIGC-as-a-service}, where AIGC service providers can deploy AI models on edge servers to provide instant services to users over wireless networks, offering a more convenient and personalized experience \cite{du2023enabling}. 
Leveraging the resource advantages at the edge enhances the accessibility of AIGC services and fosters new venues for real-time and interactive applications.
However, edge-assisted AIGC services also introduce new challenges, including optimizing resource allocation, minimizing latency, and ensuring high quality of generated contents \cite{xu2024unleashing}.
  
To accommodate various user requirements, it is essential to deploy multiple AIGC models at the edge server. While various AIGC models possess distinct capabilities, it is challenging to improve user's experience by only pursuing the high efficiency of small models or the high quality of generated contents by large models \cite{kong2023edge}. Therefore, it is crucial to efficiently assign AIGC models and allocate edge resources considering the following aspects \cite{yang2023adaptive}.
First, model assignment can be intricate because user tasks have unique performance requirements, which in turn necessitates a dynamic, task-oriented approach to assign the appropriate models \cite{xiangxiang2024axiomvision,wang2024dependence,wu2023characterizing}. Second, given the limited computing resources available on edge devices, efficient resource allocation strategies should also be investigated.  
Effective management is essential to balance the competition for computing resources among models and ensure optimal generated quality without introducing long response delay \cite{ren2022efficient}.
Many existing studies have considered the deployment of large models for distributed processing at the edge. For instance, Li \textit{et al.} harnessed the parallelism of multiple edges to improve the generation efficiency of high-resolution images according to the similarity of adjacent diffusion processes \cite{li2024distrifusion}. 
%Related study considers the problem of assigning the appropriate ASP to the users at the edge \cite{du2024diffusion}. Liu \textit{et al.} \cite{liu2024semantic} combine edge AIGC service with semantic communication to reduce transmission delay and improve the efficiency of AIGC service. However, current researches focus more on the selection and application of AIGC models at the edge, while ignoring the demand and appropriate allocation of resources among models at the edge.
%In addition, Liu \textit{et al.} \cite{liu2024semantic} combined edge AIGC services with semantic communication to improve the efficiency of AIGC services.
While current studies often emphasize training and deploying AIGC models on the edge  \cite{liu2023towards}, they tend to ignore the significant resource requirements and the need for effective resource allocation among these models, which is crucial for handling intense workloads and improving users service quality.

In this paper, we take the Text-to-Image generation task, one of the fundamental and popular AIGC services, as an example and propose an edge-enabled AIGC service system, consisting of a user prompt-based model assignment module and an adaptive resource allocation module, aiming to generate high-quality content with low latency. 
\emph{First}, we demonstrate significant differences in score distribution across various categories of prompts. %In addition, the performance of prompts with same score level on different models reveals their own scoring potential. 
Based on such observation, a model assignment method is proposed to appropriately assign models to each task. 
\emph{Second}, for the diffusion-based generation model, we conduct experiments to analyze how the performance of models and the corresponding denoising process impact both the content quality and time consumed to generate results.
\emph{Third}, an adaptive denoising step selection and computation resource allocation algorithm is designed to optimize the utilization of available resources at the edge, thereby effectively reducing time consumption and enhancing content quality. The main contributions of this paper can be summarized as follows.
\begin{itemize}
	\item We conduct experiments on a challenge category-based prompt dataset, revealing the approximate Gaussian distribution of CLIPScore for different categories, as well as the key factors influencing generation quality and time overheads for different generative models.
	\item We propose an edge-enabled AIGC service system that offloads AIGC task requests to the edge. By factoring in computational resource constraints, this system adaptively assigns models and allocates resources for high-quality content generation in response to user prompts.
	\item We design a probabilistic model assignment method and a simulated annealing-based resource allocation algorithm to solve the model assignment and computation resource optimization problem respectively, realizing the trade-off between the generation quality and the time consumption.
\end{itemize}

The remainder of this paper is organized as follows. In Section \ref{sec2}, we describe our motivations and experimental observations. In Section \ref{sec3}, we present system model and problem formulation, as well as algorithm design. Section \ref{sec4} presents the performance evaluation of the simulation results. Finally, we conclude this paper in Section \ref{sec5}.

\section{Motivation}\label{sec2}

\subsection{The Score Distribution of Prompts}
%\begin{figure}[!t]
%	\centering
%	%\vspace{-0.6cm}
%	%\noindent\hspace*{-1.3cm}
%	\includegraphics[scale=0.25]{fit curve.pdf}
%	%\vspace{-1.6cm}
%	%\captionsetup{font={small}}
%	\caption{Fitting Gaussian distribution curve of four types of prompts.}
%	\label{curve}
%	\vspace{-0.3cm}
%\end{figure}

\begin{figure}[!t]
	\vspace{-0.18in}
	\centering
	\subfigure[Basic.]{
		\includegraphics[width=0.225\textwidth]{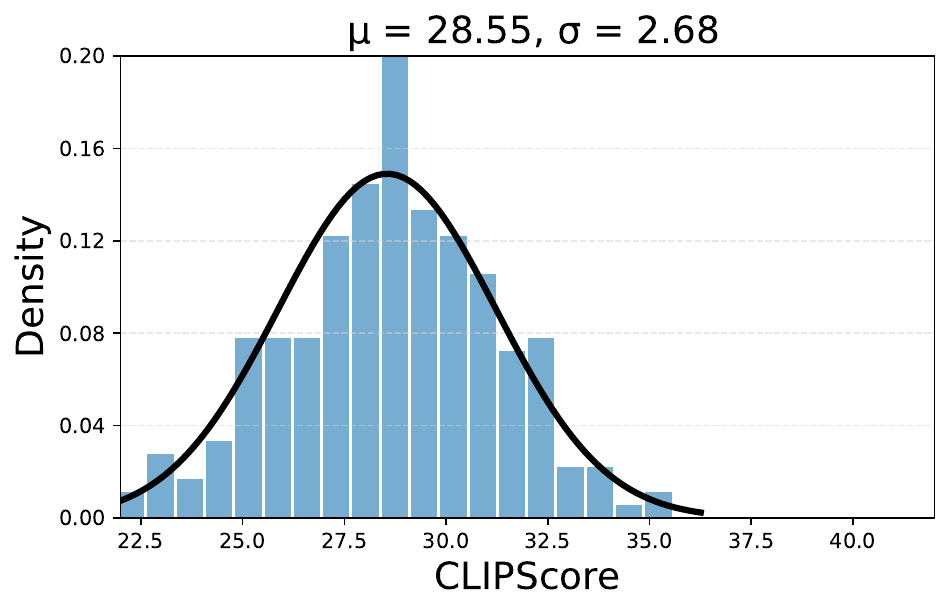}
		\label{fig1}
	}
 \vspace{-0.2cm}
        \hfill
	\subfigure[Detail.]{
		\includegraphics[width=0.225\textwidth]{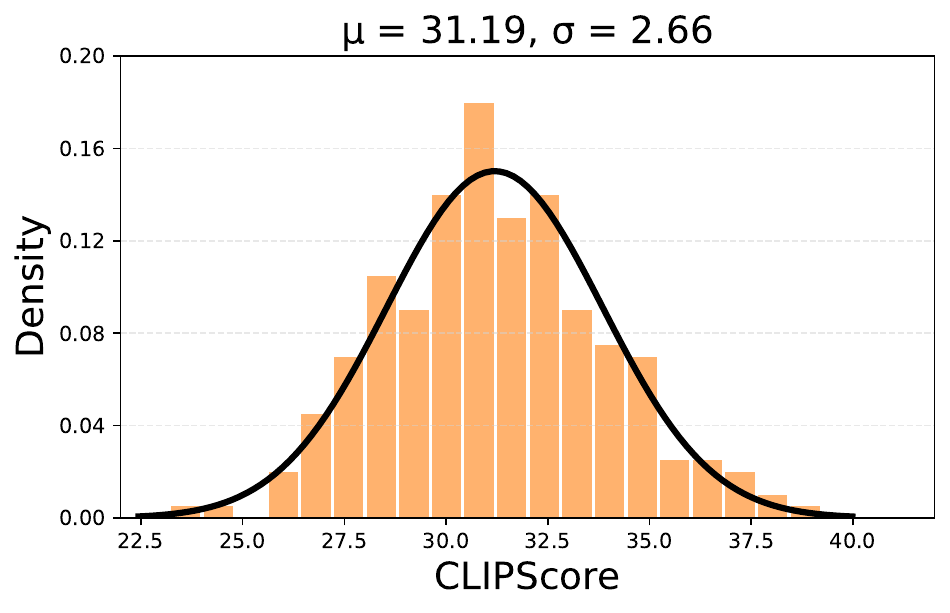}
		\label{fig2}

	}
	\hfill
	\subfigure[Imagination.]{
		\includegraphics[width=0.225\textwidth]{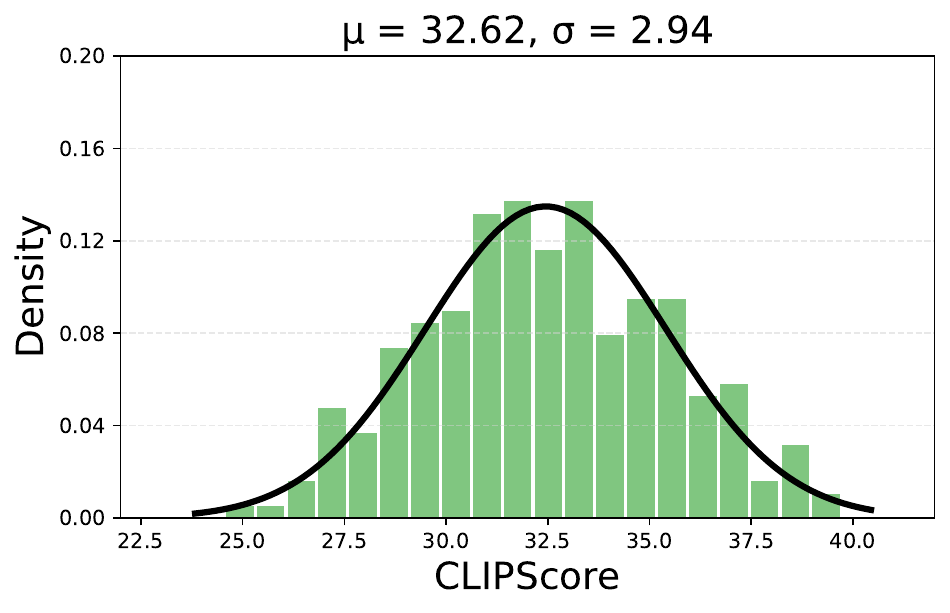}
		\label{fig3}
	}
  \vspace{-0.2cm}
	\hfill
        \subfigure[Complex.]{
		\includegraphics[width=0.225\textwidth]{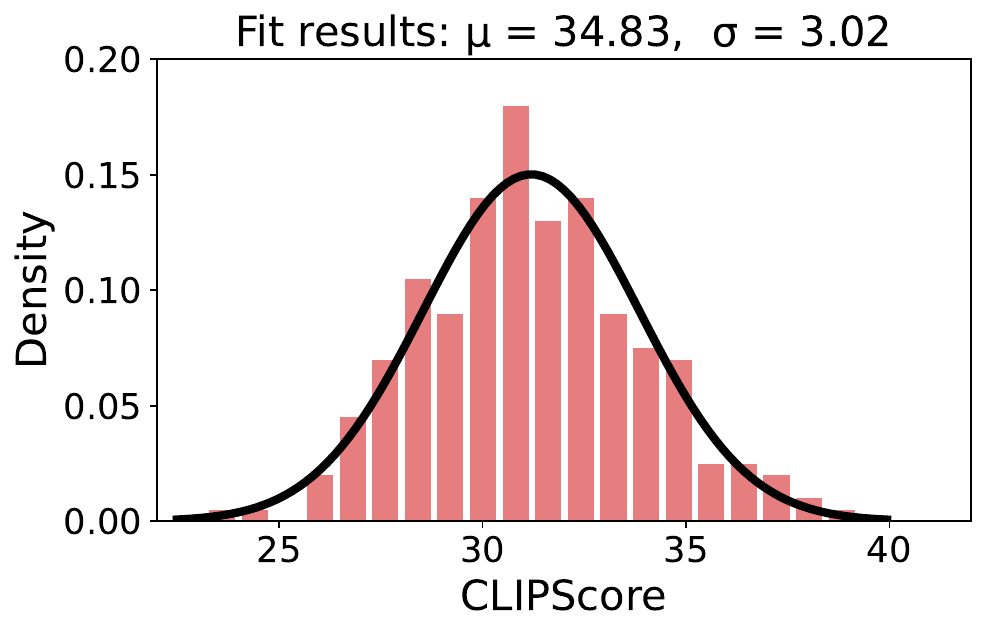}
		\label{fig4}
	}
	%\captionsetup{font={footnotesize}}
	\caption{Fitting Gaussian distribution curve of four categories of prompts.}
	\label{fit}
	\vspace{-0.25in}
\end{figure}

%Such tasks often require models to not only understand language and concepts in text, but also to be able to translate those understandings into visual representations. From creative industries to scientific research, the Text-to-Image AIGC opens up new possibilities for human creative expression and knowledge dissemination. For example, Text-to-Image AIGC can assist users in artistic creation and visualization of abstract concepts according to prompts.
%\begin{figure}[htbp]
%	\centering
%	\includegraphics[scale=0.56]{motiv_models.jpg}
%	\captionsetup{font={small}}
%	\caption{Images generated by different AIGC models when prompt is "A cinematic shot of robot with colorful feathers".}
%	\label{models}
%\end{figure}
%\begin{figure}[htbp]
%	\centering
%	\includegraphics[scale=0.62]{motiv step.jpg}
%	\captionsetup{font={small}}
%	\caption{Images generated by different denosing steps when prompt is "A cinematic shot of robot with colorful feathers".}
%	\label{steps}
%\end{figure}
 
In Text-to-Image generation tasks, the prompt refers to the textual description provided by users to generate images. Due to the differences in generation techniques and training datasets used by different AIGC models, the same prompt can generate diverse results under different AIGC models. We utilize the widely adopted CLIPScore as the metric to evaluate generative task quality. As a no-reference metric, CLIPScore builds upon Contrastive Language Image Pre-training (CLIP) model, which takes into account both the text-image alignment and image quality, showing the highest correlation with subjective human assessment \cite{hessel2021clipscore}.

In order to explore the characteristics of the same AIGC model responding to different categories of prompts, we conduct experiments on the Stable Diffusion v1.5 (SD1.5) model \cite{rombach2022high} with the PartiPrompts (P2) dataset\footnote{https://huggingface.co/datasets/nateraw/parti-prompts}. The dataset consists of over 1600 prompts, which has been labeled according to the challenge aspect. We combine prompts with similar CLIPScore in the dataset into four categories named Basic, Imagination, Detail, Complex.
%including basic, simple detail, imagination, fine-grained detail, complex, etc.
%实验中，我们统计了每一类prompt在sd1.5模型中生成结果的最优得分并观察其分布情况。

In the experiment, we calculate the CLIPScore for each category of prompts by the SD1.5 model and observe its distribution. 
The results are shown in Fig. \ref{fit}, revealing significant mean and distribution differences across CLIPScore from different categories. It can be observed that in each category, the distribution of scores approximates a Gaussian distribution.
The results enhance our understanding on the possible quality of generated contents across different categories and indicate the expected score level to be estimated for each prompt based on its category label. It also facilitates the evaluation of prompt differences for prompt-based model allocation.

\begin{figure}[!t]
        \vspace{-0.1in}
	\centering
	\includegraphics[scale=0.28]{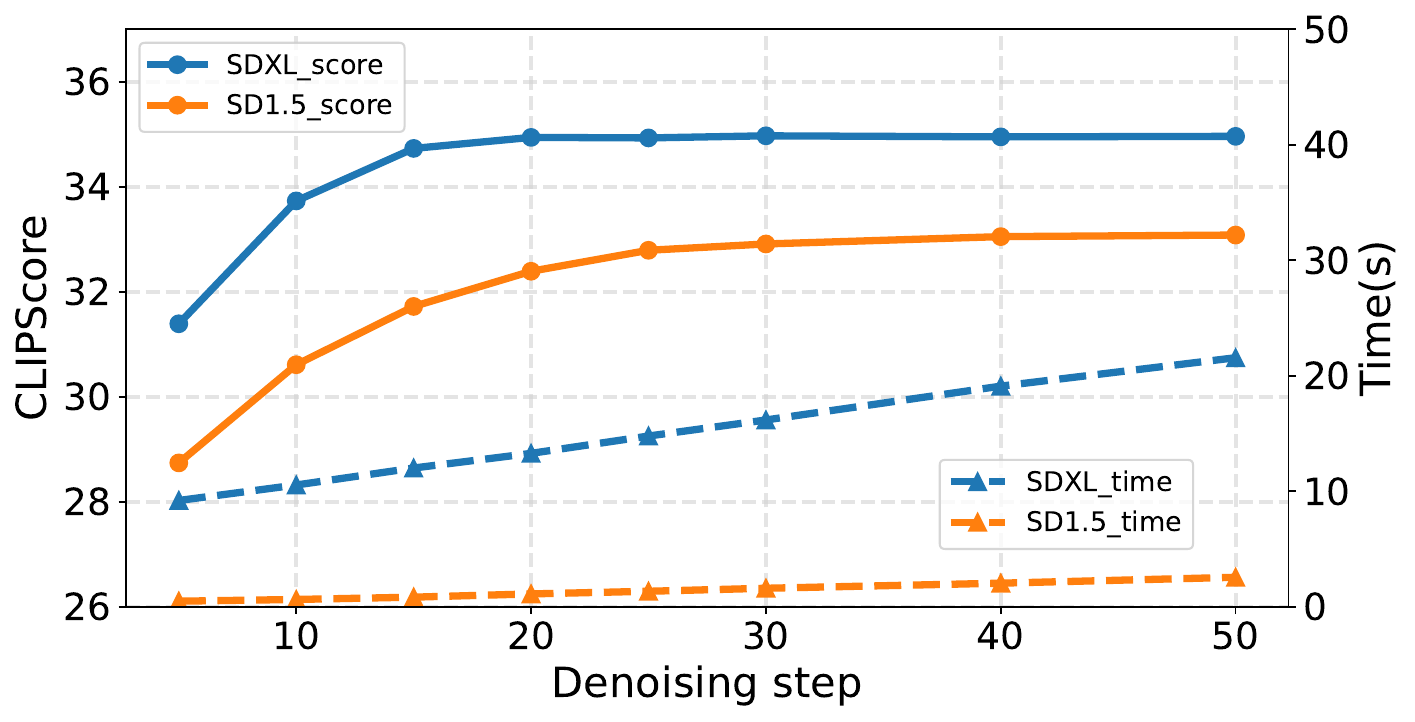}
	%\captionsetup{font={footnotesize}}
	\vspace{-0.1in}
	\caption{Models performance across different denoising steps}
	\label{Model}
	\vspace{-0.25in}
\end{figure}
\begin{figure*}[!t]
	\vspace{-0.1in}
	\centering
	\includegraphics[scale=0.68]{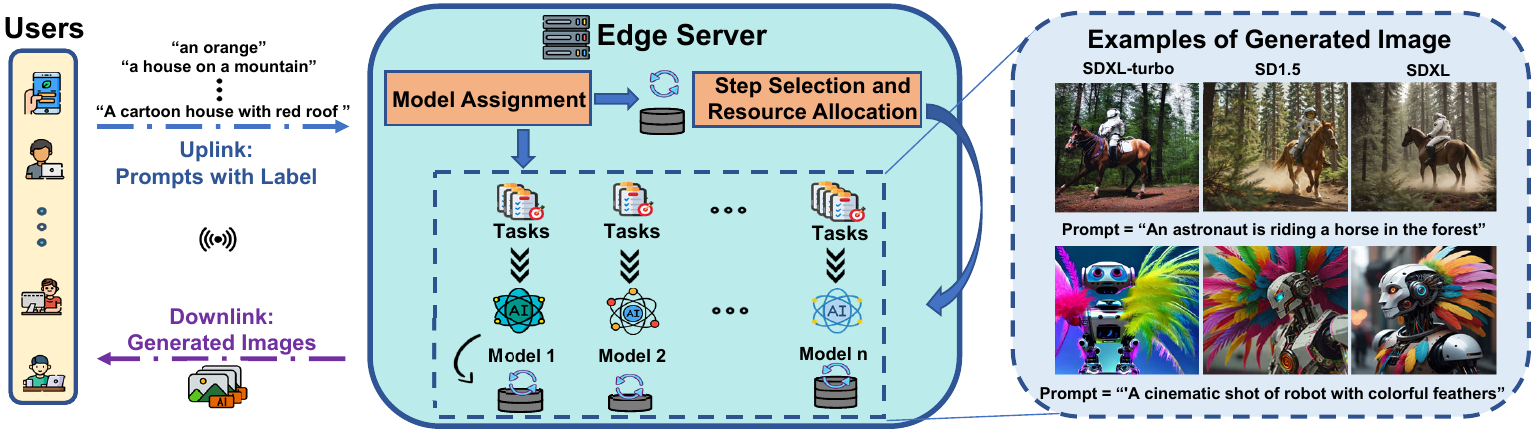}
	%\captionsetup{font={footnotesize}}
	\caption{An overview of proposed system. The examples of generated image are the result of three different models.}
	\label{system}
	\vspace{-0.2in}
\end{figure*}

\subsection{AIGC-as-a-Service on the Edge}
%Due to the high resource requirements of AIGC tasks and the limitation of user devices, it is challenging for users to perform AIGC tasks locally. Therefore, deploying the pretrained AIGC model to the edge server as a service and providing it to users through the wireless network is a promising system scheme.
To explore the feasibility of deploying AIGC models at the edge, we conduct a text-to-image experiment using the SD1.5 model on an RTX 3080Ti GPU. A total of 200 prompts are randomly selected from aforementioned P2 dataset.
For the diffusion-based AIGC models, \textit{e.g.}, SD1.5, image generation involves a gradual denoising process starting from pure Gaussian noise. Moreover, the degree of image refinement is determined by the number of denoising steps.
%For diffusion-based AIGC models, the number of denoising steps is an important parameter to control the generation process. 
Visually, increasing the number of denoising steps progressively enhances image quality, particularly in the details of the image. As shown in Fig. \ref{Model}, the generation time increases linearly with the number of steps. While the quality significantly improves up to a certain score, the marginal gain of CLIPScore diminishes as the number of steps continues to increase.
%In order to explore the feasibility of the AIGC model serving the edge, we randomly selected 200 prompts in the P2 dataset and conduct a text-to-image experiment using SD1.5 model on the RTX 3080Ti GPU. For diffusion-based AIGC model, the number of denoising steps is an important parameter. Visually, as the number of denoising steps increases, the quality of the generated image also gradually improves, especially in terms of detail. As illustrated in Fig. \ref{Model}, it takes only a few seconds for a single prompt to generate an image. In addition, the time consumption shows a linear increase with the number of denoising steps increasing, while the generation quality significantly improves within a certain number of steps. Then, the marginal gain decreases as the number continues to increase.

Note that, AIGC models are rapidly evolving, higher quality images often generated at the cost of higher computation resource requirements. Therefore, we conduct the same experiment on the upgraded SDXL model \cite{podell2023sdxl}. As shown in Fig. \ref{Model}, with the same prompts, the SDXL model has a substantial improvement in CLIPScore compared with SD1.5, but at much higher time consumption.
This is due to the fact that the number of parameters of SDXL is 3.5 billion (B), which is over three times that of SD1.5 with parameters being 1.06B. Besides, it takes 338 TFLOPs for SDXL to generate a 1024*1024 image with the denoising steps of 25. The result demonstrates that more model parameters lead to better performance but also a significant increase in computation resources demand \cite{li2024distrifusion}. 
% This is due to the fact that the number of parameters of SDXL is 3.5 billion (B) compared to 1.06B of SD1.5.
%During the user's interaction with the edge AIGC service, they need to wait for the edge AIGC model to complete the generation tasks and return the results.
While image quality is crucial, system latency also significantly impacts the experience of users.
Our experiments demonstrate that the number of denoising steps is a critical resource in generative models, influencing both generation quality and time consumption.
Therefore, at the edge where multiple models are deployed, the number of denoising steps should be properly selected. The corresponding computing resources should also carefully allocated to generate high quality images with low latency. 

\section{System Model and Problem Formulation}\label{sec3}

\subsection{System Overview}

Consider a scenario where a set of mobile user simultaneously send requests to the same edge server for AIGC services. As shown in Fig. \ref{system}, a set of generative models are deployed to effectively meet the users' diverse requirements.
%Consider a scenario where multiple users connecting to the same edge server request AIGC services on their own end devices. Since the user device is usually too resource-limited to process the AIGC tasks, it is necessary to wisely schedule the available resource on the edge server to complete AIGC services. As different AIGC models exhibit different generation abilities and time consumptions, we deploy multiple models to meet the user requirements. 
%Our proposed edge-enabled AIGC service system consists of two components: \textit{probabilistic model assignment} and  \textit{adaptive denoising step and computing resource allocation}. 
After receiving the requests, the edge assigns the model for each prompt of the user. Then, with selected denoising steps, generation process is performed on the edge with allocation of computing resources for each model. Finally, the generated contents are returned to the mobile user.
Considering the unknown content quality generated by the AIGC models and the prior knowledge of the score distribution from different categories of prompts, we adopt a \textit{probabilistic model assignment} strategy to efficiently assign the most suitable model for each user. Then, we design an \textit{adaptive denoising step and computation resource allocation} algorithm to efficiently utilize the computation resources on the edge server.
\subsection{Score and Latency Model}
The set of models deployed on edge, $\mathcal{M} = \{1,\dots, m,\dots,M\}$, is considered to have varying generation abilities. The users send prompt requests with category labels through the device, which is defined as the task set $\mathcal{N} = \{1,\dots, n,\dots,N\}$. Without loss of generality, task requests arrive at the edge at the beginning of each time slot.
%and the number of task requests $\mathcal{N}$ in each time slot follows the Poisson distribution in the whole time period. The number of tasks accepted by each model is $N_{i}^t$.
\subsubsection{\textbf{Generated score model}}
As observed in Section II, the denoising steps, $s_m$, has a crucial impact on the quality of the generated image (\textit{i.e.}, CLIPScore). $C_m^n$ is denoted as the CLIPScore for the $n$-th task of processing $s_m$ steps by model $m$. We introduce a binary variable, $x_m^n$, to indicate whether model $m$ is selected by the $n$-th task. Therefore, the CLIPScore for the $n$-th task can be represented as:
\begin{equation}
	C_n=\sum_{m=1}^{M} x_m^n C_m^n(s_m).
\end{equation}
\subsubsection{\textbf{Latency model}}
For an AIGC service request, the system delay includes prompt transmission delay, $d_{1}$, and AIGC model inference delay, $d_{2}(s_{m},\gamma_{m})$, which is determined by the selected steps and the resources allocation. Specifically,
\begin{equation}
	d_{2}(s_{m},\gamma_{m})=\alpha_m \cdot d^{\ast}(s_m),
\end{equation}
where $\alpha_m=\frac{\mathit{\Gamma}}{\gamma_m}$ denotes the ratio of the total computation resource $\mathit{\Gamma}$, compared to the resource allocated $\gamma_m$ by the model in TFLOPS. $d^{\ast}(s_m)$ represents the relationship between the number of denoising steps and inference latency using available resources $\mathit{\Gamma}$. The transmission time of the generated image from the model back to the user is denoted as  $d_3$. Then, the overall delay of the $n$-th task can be formulated by:
\begin{equation}
	D_m^n(s_m,\gamma_{m})=d_1+d_{2}(s_{m},\gamma_{m})+d_{3}.
\end{equation}

As the time consumption of AIGC model inference $d_2$ is significantly larger than the two transmission delays, $d_1$ and $d_3$ are thus ignored in our system.
\subsection{Problem Formulation}
In our edge-enabled AIGC service system, each prompt requested from the user needs to be assigned with a model $m$ for content generation. Besides, at the beginning of each time slot, multiple models run simultaneously to take advantage of all available computation resources. Therefore, the computation resources $\mathit{\Gamma}$ and the appropriate number of denoising steps $s_{m}$ should be carefully determined for each model. In order to achieve a better trade-off between generation quality and time consumption for the tasks handled by each model, we design the following utility function and aim to maximize the overall utilities under limited edge computation resources. The problem can be formulated as follows:
\begin{align}
	\mathcal{P}:\quad \max\limits_{\boldsymbol{x},\boldsymbol{s},\boldsymbol{\gamma}} \frac{1}{N}\sum_{n=1}^{N} & \sum_{m=1}^{M} x_m^n [C_m^n(s_m) - \omega \cdot   D_m^n(s_m,\gamma_m)] \\
	\text{s.t.} \quad C_{1}&: \sum_{m=1}^{M}x_m^n=1 ,  \forall n\in \mathcal{N},\\
 	C_{2}&: x_m^n \in \{0,1\} , \forall m\in \mathcal{M} , \forall n\in \mathcal{N}, \\
	C_{3}&: s_m \in \mathcal{S} , \forall m\in \mathcal{M},  \\
	C_4&: \sum_{m=1}^{M} \gamma_m \le \mathit{\Gamma},
\end{align}
%$$\quad C_5: \gamma_i \ge 0 ,\forall i\in M$$
where the set $\mathcal{S} = \{s^1,s^2,\dots,s^k\}$ denotes the optional denoising steps. The weight $\omega$ controls the trade-off between CLIPScore and time consumption. Therefore, the optimal solution of the problem $\mathcal{P}$ trades the average score for lowering the time consumption. $C_1$ and $C_2$ ensure that, one and only one model is selected to process each prompt task. $C_3$ indicates the range of selection for denoising steps. $C_4$ ensures that the computation resources allocated by the models do not exceed the available computation resource $\mathit{\Gamma}$ at the edge.

Due to the optimization problem $\mathcal{P}$ is a mixed integer non-linear programming (MINP), it is NP-hard \cite{xu2018joint}. Therefore, we decouple the problem into two subproblems. We first address the model assignment problem, and then the problem of step selection and computation resource allocation is solve with approximate optimality.

\begin{figure}[!t]
	\centering
	\begin{minipage}[t]{0.5\textwidth}
		\centering
		\includegraphics[scale=0.32]{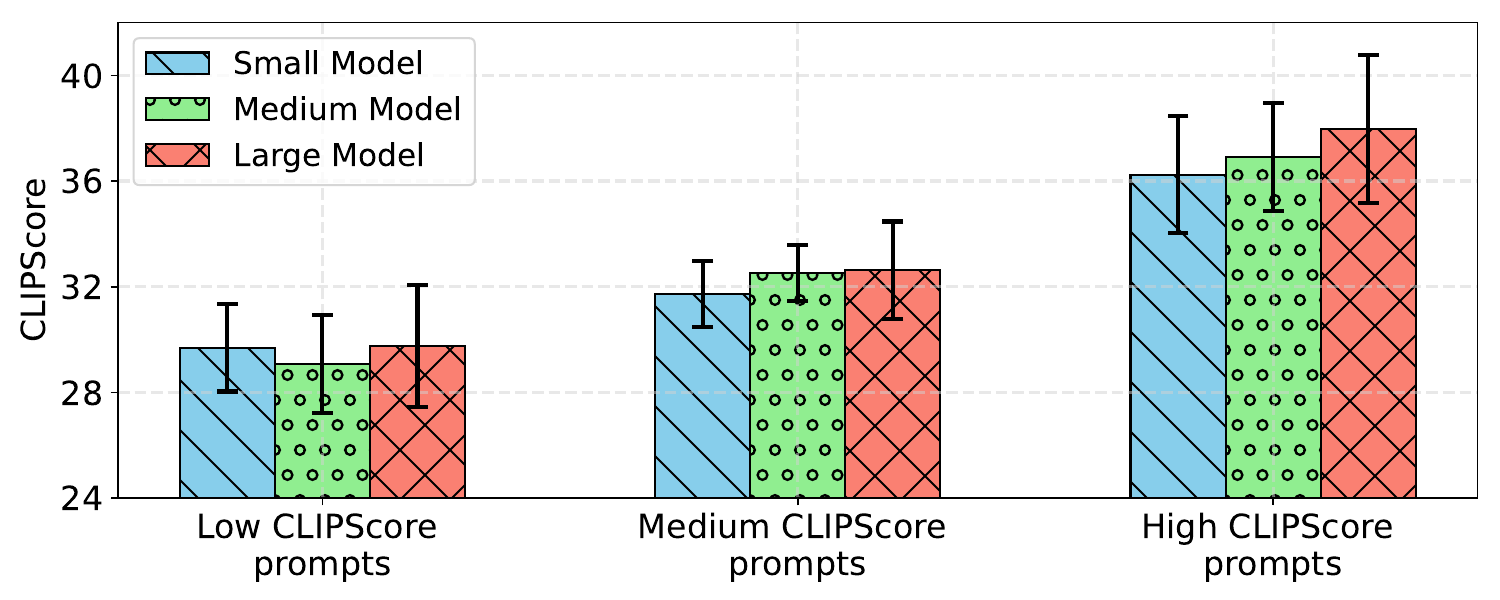}
		%\captionsetup{font={footnotesize}}
		\vspace{-0.2cm}
		\caption{CLIPScore of three different score level prompts with different models.}
		\label{motiv}
	\end{minipage}
	\vspace{-0.4cm}
\end{figure}

\subsection{Probabilistic Model Assignment}
According to the analysis in Section \ref{sec2}, even prompts within the same category exhibit differences in CLIPScore, and different models distinctly affect both the generation quality and time consumption. Therefore, we propose adopting specific model assignment methods to enhance the overall utility of the tasks. 
We take three models with different generation capabilities and computational costs as examples, named small, medium, and large model, where the larger model generates better results with longer delay. Then, we set two thresholds, dividing the CLIPScore into three intervals. Prompts that achieve different scores on the medium model are reclassified according to the score interval into three score levels. These are then processed separately on the small and large models, with the result presented in Fig. \ref{motiv}.
We observe that prompts with low CLIPScore on the medium model can save time on the small model and do not cause the loss of average scores. Besides, prompts that score high on the medium model have a higher cost of time on the large model, but in turn get a significant CLIPScore gain. %This can also be seen as prompt's own scoring potential. 

In real-world applications, it is difficult to accurately profile the score level for prompts of different categories with each model, especially with those that require huge computation overhead. Therefore, given the distribution curves of prompt CLIPScore for different categories on the medium model, we consider using the expected score probability for the model assignment of different categories.
For example, when the score of a prompt is expected to be lower than $x_1$ on the medium model, it is considered as a low score level prompt and will be assigned to the small model with greater probability. 
In other words, according to the defined score intervals, the probability of each prompt category being assigned to each model is calculated as the cumulative probability density of its score distribution on the medium model within each interval.
%In other words, the probability of a prompt scoring within three defined intervals determines its assignment to one of the three models. 
%In other words, the probability that a category of prompt score in three intervals is the probability that the category of prompt is assigned to three models. Therefore, depending on the score probability distribution results from different categories of prompts on the medium model, each category of prompt is assigned to the small model with a probability of scoring less than $x_1$ and to the large model with a probability of scoring higher than $x_2$.
%So, we set the score threshold, so that the category with low score level has a greater probability of being assigned to the light model, the category with high score level has a greater probability of being assigned to the large model, and the remaining two categories are more likely to be assigned to the medium model.
%According to the different performance of the three models and the required computing power, 
% we can see that for prompt categories such as basic with low scores, even if more denoising steps are invested in using a better model, the score gain of the generated images is still low. In contrast, categories with higher scores gain more when processed on a better model.

Specifically, consider the assignment strategy when there are three performance models $m\in{[1,2,3]}$. The prior score distribution of four categories $A_{j\in{[1,2,3,4]}}$ on the medium model 2 follows Gaussian distribution, and the mean $\mu_{j}$ and standard deviation $\sigma_{j}$ are known in advance. Therefore, the probability $P_j^m$, that category $A_j$ is assigned to model $m$ can be calculated according to:
\begin{equation}
	P_j^m = \int_{h_1}^{h_2}{\frac{1}{\sqrt{2\pi \sigma_{j}^{2}}}e^{-\frac{(x-\mu_{j})^2}{2\sigma_{j}^{2}}}}dx,\\
\end{equation}
where $h_1$ and $h_2$ are defined on different models as:
\begin{equation}
h_1,h_2=\begin{cases}
		0, x_1   .  \quad m=1\\
		x_1, x_2 .  \quad m=2 \\
		x_2, \infty . \quad m=3
	\end{cases}.
\end{equation}
\begin{figure}[!t]
	\vspace{-0.12in}
	\renewcommand{\algorithmicensure}{\textbf{Output:}}
	\begin{algorithm}[H]
		\caption{Step Selection and Resource Allocation }
		\label{resource allocation}
		\begin{algorithmic}[1]
			\Require $N_m$, $\mathcal{M}$, function $\hat{C_m}$, $D_m$, $\mathcal{S}$, $\gamma$
			\Ensure $S^{\ast}=[s_1, s_2, s_3]$, $\mathit{\Gamma}^{\ast}=[\gamma_1, \gamma_2, \gamma_3]$\\
			Initial $S^{\ast}$, $\mathit{\Gamma}^{\ast}$, temperature $T$ and cooling coefficient $\beta$\\
			$U(S^{\ast}, \mathit{\Gamma}^{\ast}) \gets $ compute initial utility 
			\While{$T > T_{min}$}
			\For{$k=1$ to $k_m$}
			\State $\tilde{S}$, $\tilde{\mathit{\Gamma}} \gets $ generate a new candidate solution 
			\If{$GetLatency( \tilde{S}, \tilde{\mathit{\Gamma}}) < t$}
			\State $U(\tilde{S}, \tilde{\mathit{\Gamma}}) \gets $ compute new utility 
			\State $\Delta U= U(\tilde{S}, \tilde{\mathit{\Gamma}}) - U(S^{\ast}, \mathit{\Gamma}^{\ast})$
			\If{$\Delta U > 0$}
			\State $S^{\ast}, \mathit{\Gamma}^{\ast} \gets $ accept new solution
			\Else
			\State $p \gets $ generate random number in [0,1]
			\If{$p < \exp(\Delta U / T)$}
			\State $S^{\ast}, \mathit{\Gamma}^{\ast} \gets $ accept new solution
			\EndIf
			\EndIf					
			\EndIf
			\State $T \gets T \times \beta$
			\EndFor
			\EndWhile
			\State \Return $S^{\ast}$, $\mathit{\Gamma}^{\ast}$
		\end{algorithmic}
	\end{algorithm}
	\vspace{-0.3in}
\end{figure}

\subsection{Step Selection and Resource Allocation}
%Instead of selecting steps individually for each task because the scoring trends are consistent across tasks relative to the same model, we consider assigning the same number of steps and computing resources to each model based on the difference in the number of tasks assigned to each model.
Instead of selecting steps individually for each task, we consider allocating the same number of steps $s_m$ and computational resources $\gamma_m$ to each model based on the difference in the number of tasks assigned to each model. In order to solve this problem, we propose a step selection and resource allocation algorithm based on simulated annealing (SA), aiming to efficiently search for sub-optimal solutions. As presented in Algorithm 1, it iteratively seeks better solutions by setting policies to modify the current solution, and accepts them only if they improve or meet the Metropolis criteria. The utility function in our algorithm is denoted as follows:
\begin{equation}
	U= \frac{1}{N} \sum_{n=1}^{N} [\hat{C_m^n}(s_{m})-\omega \cdot D_m^n(s_{m},\gamma_{m})], %- \omega_2 \cdot E
\end{equation}
where $\hat{C_m^n}(s_{m})$ is the estimated score function. Among them, $\hat{C_m^n}(s_{m})$ and $D_m^n(s_{m},\gamma_{m})$ are different for different models. The temperature parameter gradually decreases until the termination condition is reached, and the rate of decrease is related to the cooling coefficient $\beta$, where $\beta \in (0,1)$. While satisfying constraints $C_{1-4}$, algorithm \ref{resource allocation} solves the problem $\mathcal{P}$ with near-optimal solution of steps and resources.
\begin{figure*}[ht]
	\vspace{-0.4cm}
	\centering
	\subfigure[The performance of three allocation methods.]{
		\includegraphics[width=0.30\textwidth]{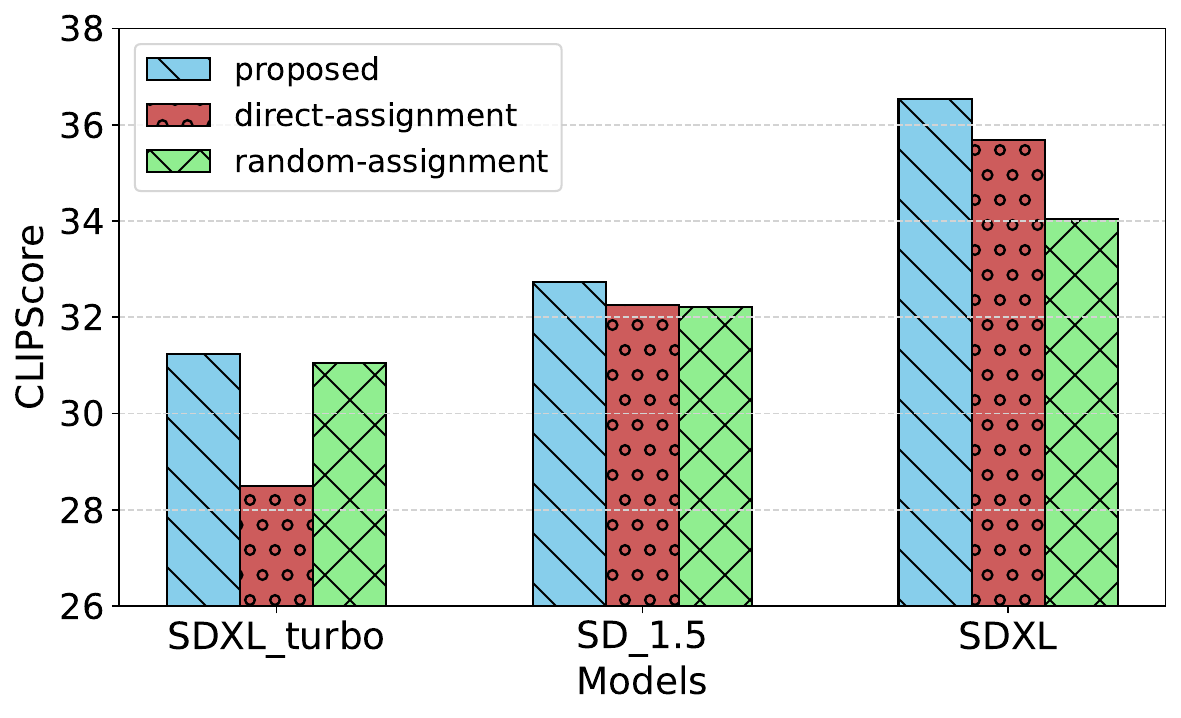}
		\label{fig:sub1}
	}
	\hfill
	\subfigure[Average utility under different time slots.]{
		\includegraphics[width=0.30\textwidth]{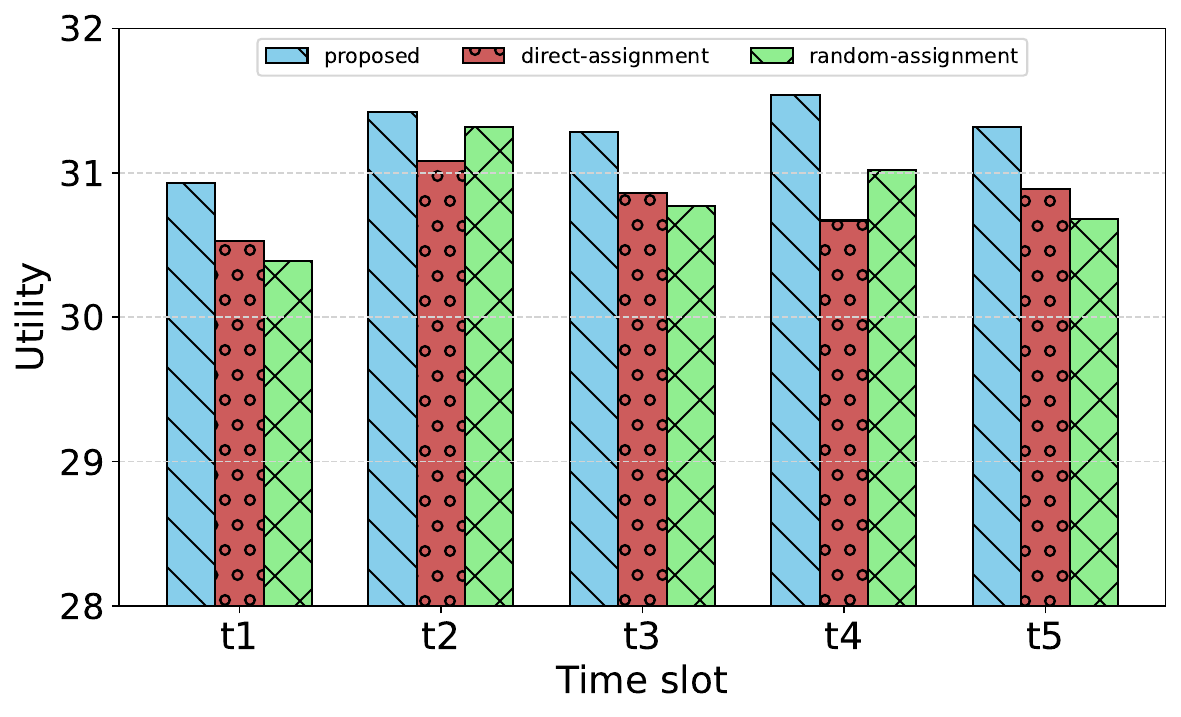}
		\label{fig:sub2}
	}
	\hfill
	\subfigure[CLIPScore of three allocation methods.]{
		\includegraphics[width=0.305\textwidth]{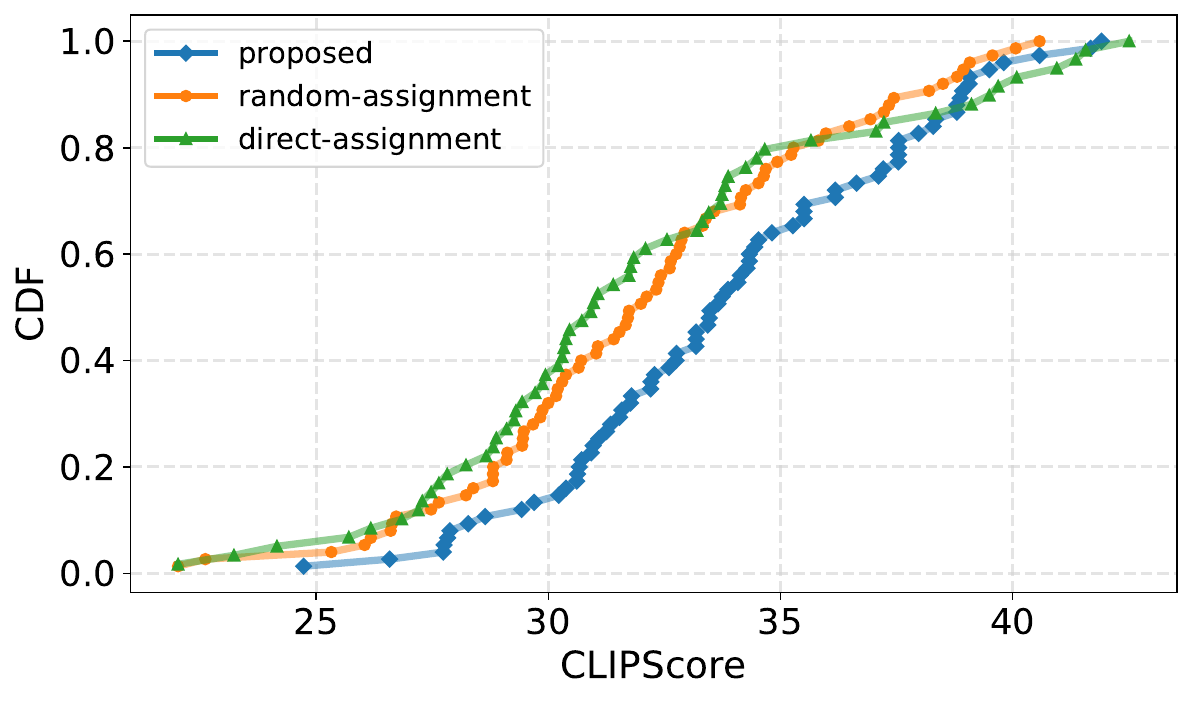}
		\label{fig:sub3}
	}
	\hfill
	%\captionsetup{font={footnotesize}}
        \vspace{-0.1in}
	\caption{  The performance comparison of different model assignment methods.}
	\label{fig:test}
	\vspace{-0.15in}
\end{figure*}

\begin{figure*}[htbp]
	\begin{minipage}[t]{0.24\textwidth}
		\centering
		\includegraphics[scale=0.27]{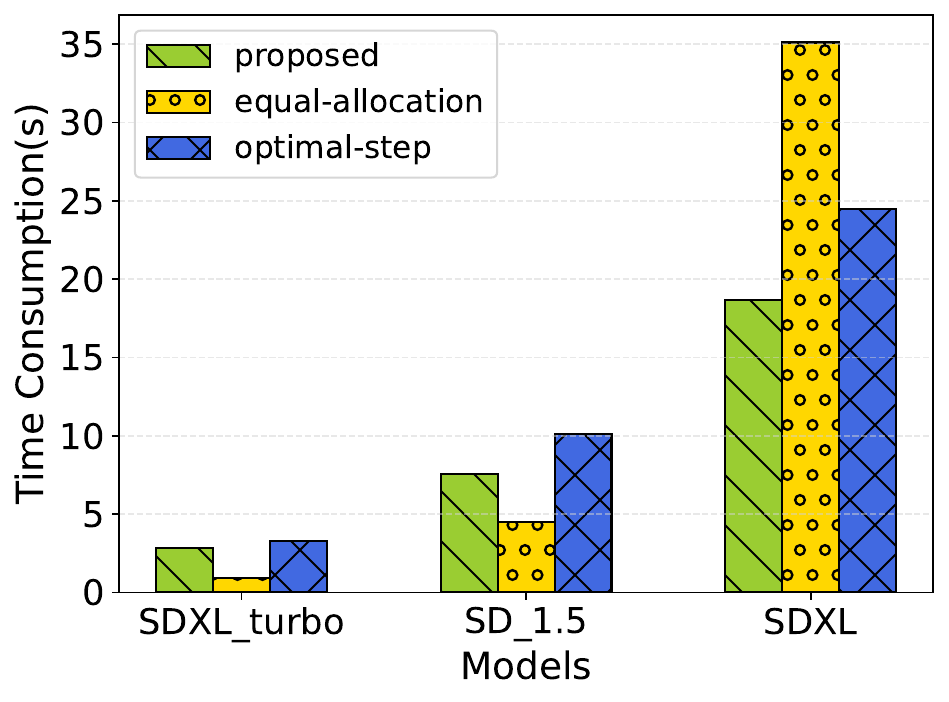}
		%\captionsetup{font={footnotesize}}
		\vspace{-0.3in}
		\caption{Average time consumption on each model.}
		\label{time}
	\end{minipage}
	\hfill
	\begin{minipage}[t]{0.24\textwidth}
		\centering
		\includegraphics[scale=0.27]{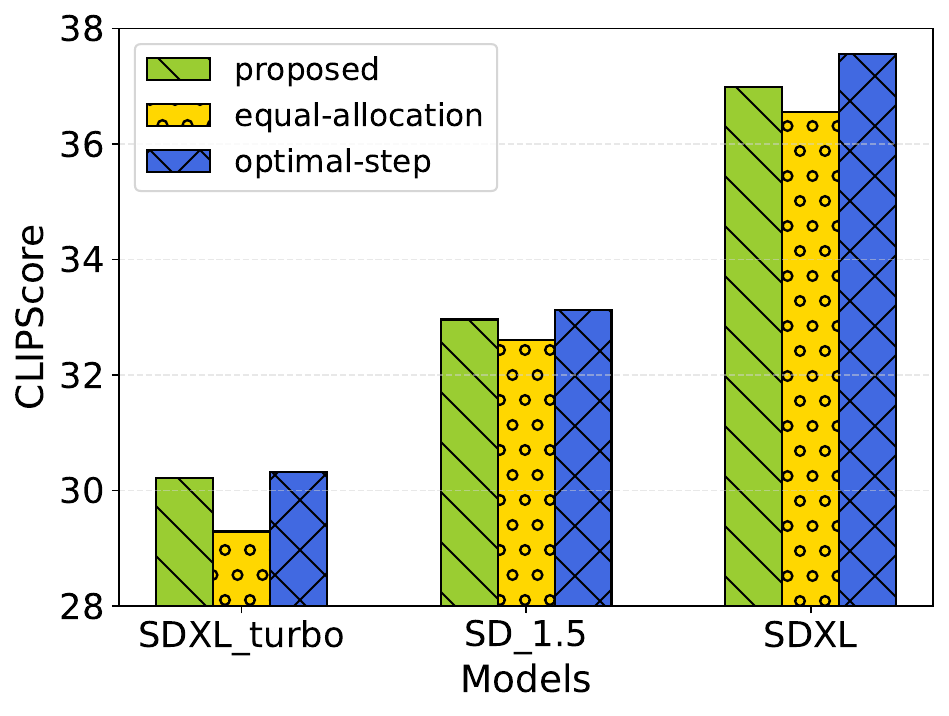}
		%\captionsetup{font={footnotesize}}
		\vspace{-0.3in}
		\caption{Average CLIPScore on each model.}
		\label{score}
	\end{minipage}
	\hfill
	\begin{minipage}[t]{0.24\textwidth}
		\centering
		\includegraphics[scale=0.27]{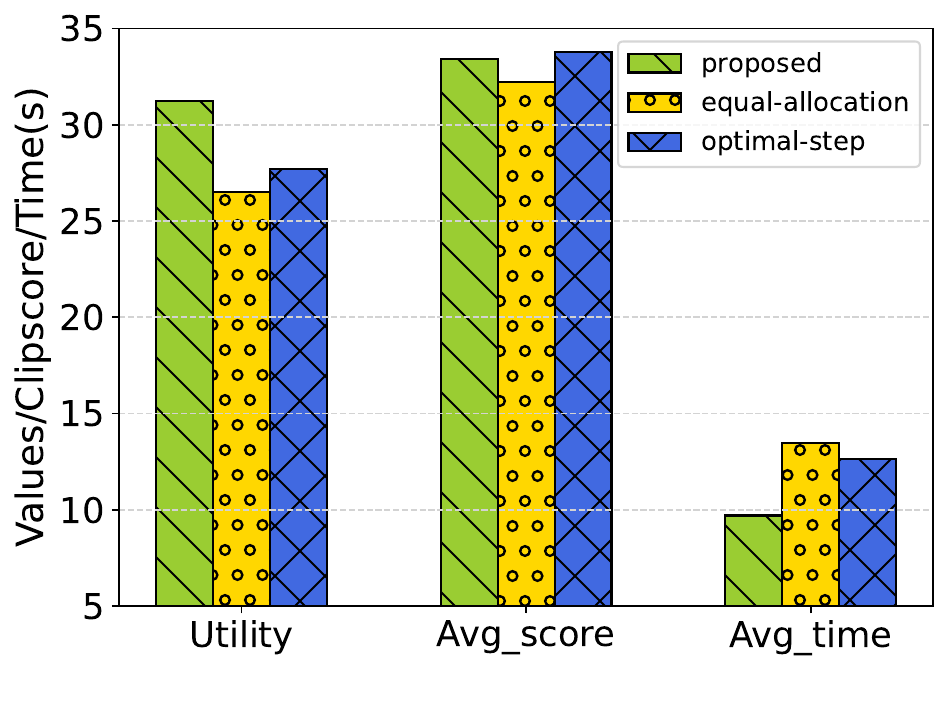}
		%\captionsetup{font={footnotesize}}
		\vspace{-0.3in}
		\caption{The overall performance comparison of the three methods.}
		\label{avgvalues}
	\end{minipage}
        \hfill
	\begin{minipage}[t]{0.24\textwidth}
		\centering
		\includegraphics[scale=0.27]{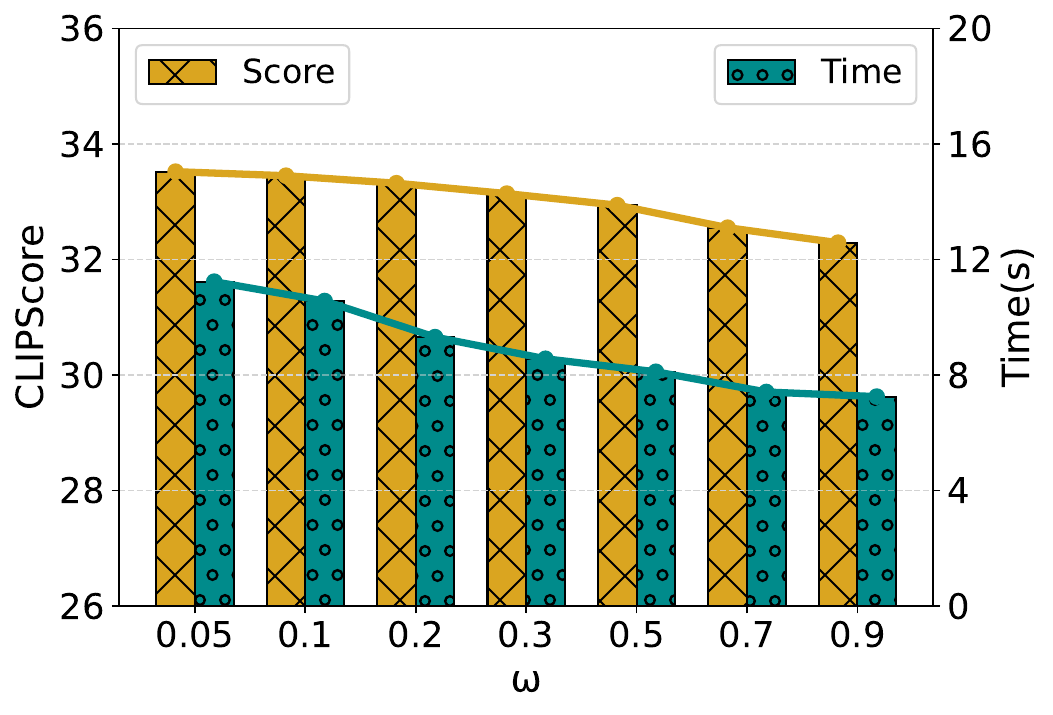}
		%\captionsetup{font={footnotesize}}
		\vspace{-0.3in}
		\caption{Impact of weight $\omega$ on performance.}
		\label{w}
	\end{minipage}
	\vspace{-0.2in}
\end{figure*}

\section{Performance Evaluation}\label{sec4}
In this section, we evaluate the performance of our proposed system. The experimental settings and results are as follows.
\subsection{Experimental Setting}
In the experiment, we take the Text-to-Image generation task as example. Besides, the edge server is deployed with models featuring varying generation abilities. 
Specifically, three models from the Stable Diffusion series are selected, \textit{i.e.}, SDXL-Turbo \cite{sauer2023adversarial}, SD1.5, and SDXL. While these models share similar structures, they exhibit different trade-offs between the quality and time required for generating results, primarily due to the variations in their training datasets and amount of parameters.
We conduct experiments using P2 dataset, the details of which have been presented in Section \ref{sec2}.
%Specifically, we choose three popular models in Stable Diffusion series, SDXL-turbo \cite{sauer2023adversarial}, SD1.5, and SDXL, which have different tradeoffs in quality and time consumption of the generated result.
%We implement experiments using P2 dataset, whose prompts is divided into four categories as mentioned in Section \ref{sec2}.
%prompts are manually marked as different challenge labels. In our experiment, 
%\textit{Parameter}:

Moreover, the AIGC model for receiving prompts and generating images is deployed on the NVIDIA GeForce RTX 3080 Ti GPU as the edge server. The utility function weight $\omega$ is set to 0.2. The CLIPScore thresholds $x_1$ and $x_2$ are set to 29.5 and 33.8. The step set is $\{$10, 14, 18, 22, 26, 30, 34, 38, 42$\}$. In particular, since the small model SDXL-Turbo is capable of producing good results in 1-4 steps, we consider fixing its denoising step number to 1. We compare our system with the following strategies:

%Compare the probabilistic model assignment of tasks with following baselines:
\begin{itemize}
	\item \textbf{Direct-assignment:} This method assigns prompts of the same category to a single model, regardless of score differences within that category.
	\item \textbf{Random-assignment:} This method randomly assigns each prompt to one of three models, giving each category an equal chance of selection.
	\item \textbf{Equal-allocation}: In this method, all the computation resources of the edge server are equally allocated to each model, and only the number of denoising steps is selected.
	\item \textbf{Optimal-step}: The method selects the near-optimal step number for each model to guarantee the highest score, while only the computation resources are allocated.
\end{itemize}

\subsection{Performance of Probabilistic Assignment}
We compare the proposed  \textit{Probabilistic-assignment} with \textit{Direct-assignment} and \textit{Random-assignment}. The average CLIPScore generated on each model is shown in Fig. \ref{fig:sub1}. Compared with two methods, our proposed method improves CLIPScore from 0.2 to 2.4, especially in SDXL model. In particular, although the \textit{Random-assignment} assigns more high score level prompts to the small model, which increases the average score of the tasks handled by the small model, it also significantly decreases its overall score on the large model. Conversely, the \textit{Direct-assignment} overlooks the variation in score levels within the same prompt category, thus limiting the scoring potential across different models, resulting in its score not being improved.
Our method utilizes the scoring potential of each prompt itself, effectively assigning each prompt to the appropriate model based on the scoring probability of different categories.
As shown in Fig. \ref{fig:sub2}, our method demonstrates an improvement in overall utility, achieving up to a 2.5\% increase compared with other methods at different time slots. It indicates that our method achieves a trade-off between score quality and time consumption. Apart from that, the task CLIPScore CDF of the three methods is shown in Fig. \ref{fig:sub3}. It can be seen that the proposed method improves the overall score level, which indicates that \textit{Probabilistic-assignment} strategy is beneficial to obtain higher scores.

\subsection{Performance of Step Selection and Resource Allocation}

We then compare the time consumption and average CLIPScore of our proposed \textit{step selection and resource allocation} method with \textit{Equal-allocation} and \textit{Optimal-step}. As shown in Fig. \ref{time}, compared with the \textit{Optimal-step}, the proposed method achieves less time consumption on each model, saving an average of 30.2\% in time consumption. This is due to the fact that the \textit{Optimal-step} method selects more denoising steps to achieve higher quality results at the cost of greater time overhead.
 In addition, the \textit{Equal-allocation} shows lower time consumption for the small model. However, this is due to the lack of consideration given to the varying resource requirements of different models. As a result, the larger models experience significantly higher time consumption. Our method saves 39.1\% time consumption compared to \textit{Equal-allocation}.

In terms of CLIPScore in Fig. \ref{score}, our proposed method approaches the \textit{Optimal-step} while improving over the \textit{Equal-allocation} on all models. Moreover, as shown in Fig. \ref{avgvalues} the proposed method respectively improves the overall utility by 17.8\% and 12.7\% compared with the other two methods. The results demonstrate that the proposed method realizes reasonable adaptive allocation of steps and resources, and obtains a balance between generation quality and time consumption.

We show the influence of weight coefficient $\omega$ in the target utility function on the overall performance in Fig. \ref{w}. The results show that when $\omega$ is small, the system will sacrifice time by choosing more denoising steps to improve the overall CLIPScore. As $\omega$ increases, the system pays more attention to the cost of time consumption.

\section{Conclusion}\label{sec5}

In this paper, we have proposed an edge-enabled AIGC service system that offloads intensive AIGC tasks to the edge for model assignment and resource allocation to provide users with efficient and personalized AIGC services.
The system performs a probabilistic model assignment method to assign appropriate models to different categories of prompts. Then, it adaptively selects the number of denoising steps and allocates computation resources for models based on the task requests received by each AIGC model.
Simulation results have demonstrated that the proposed system effectively optimizes resource allocation, achieving the trade-off between generation quality and time consumption across different performance models.
For future work, we will explore the AIGC service system for multiple types of content generation applications.

\section*{Acknowledgement}
The work was supported in part by the Natural Science Foundation of China under Grant 62001180, and in part by the Young Elite Scientists Sponsorship Program by CAST under Grant 2022QNRC001.

\footnotesize
\bibliographystyle{ieeetr}
\bibliography{ref}

\end{document}